\documentclass[12pt]{article}
\usepackage{amsmath}
\usepackage{amsfonts, color}
\usepackage{amssymb}
\usepackage{braket}
\begin{document}
\author{C. Athorne\\School of Mathematics and Statistics\\ University of
  Glasgow, University Place, Glasgow G12 8QQ, UK\\
email:Christopher.Athorne@gla.ac.uk \\ \\
L. Cruzeiro\\CCMAR/CIMAR - Centro de Ciências do Mar, and Physics\\
  FCT, Universidade do Algarve, Campus de Gambelas, 8005-139,\\ Faro, Portugal\\
  email: lhansson@ualg.pt\\
  \\ J. C. Eilbeck\\Maxwell Institute and School of Mathematical and
  Computer Sciences\\ Heriot-Watt University, Edinburgh EH14 4AS, UK\\
  email: J.C.Eilbeck@hw.ac.uk}

\title{Exact analytic multi-quanta states of the Davydov Dimer}

\maketitle
\begin{abstract}
  The Davydov model describes amide I energy transfer in proteins
  without dispersion or dissipation. In spite of five decades of
  study, there are few exact analytical results, especially for the
  discrete version of this model. Here we develop two methods to
  determine the exact orthonormal, multi-quanta, eigenstates of the
  Davydov dimer. The first method involves the integration of a system
  of ordinary differential equations and the second method applies
  purely algebraic methods to this problem. We obtain the general
  expression of the eigenvalues for any number of quanta and also, as
  examples, apply the methods to the detailed derivation of the
  eigenvectors for one to four quanta, plus a brief example in the
  case of $n=5$ and $n=6$.
\end{abstract}

\section{Introduction}





The energy for most of the processes that keep cells alive comes from
the chemical reaction of hydrolysis of adenosinetriphosphate
(ATP). This chemical reaction is catalyzed by proteins and is used for
processes like the transport of ions and other ligands across cell
membranes, protein synthesis and folding, cell division, and many
others. In spite of all the progress made about the molecular aspects
of those processes, it is not yet known how the energy released in the
hydrolysis of ATP is ultimately used for work. The basic assumption of
the model due to the Ukrainian physicist Davydov \cite{Dav1,Dav2} is
that, in proteins, the energy released is stored in the form of amide
I vibrations which are essentially stretching vibrations of the C=O
groups. Formally, the Davydov model is analogous to the polaron model
in which a free electron attracts the positive crystal sites, leading
to a local distortion which, in turn, lowers the energy of the
electron. The electron and its associated lattice distortion then move
in a correlated manner in crystal, constituting a so-called
self-trapped state or polaron.  In the Davydov model, the role of the
electron is played by the amide I vibration, and its interaction with
the lattice sites is due to the dependence of the amide I energy on
the length of the hydrogen bond that connects the C=O groups. When the
energy of the distortion of the hydrogen-bonded lattice is smaller than
the reduction in the amide I energy, the amide I also becomes
self-trapped, in a state known as the Davydov soliton
\cite{Dav1,Dav2,Scott85,Scott92}.
\\

In the continuum approximation, it is possible to obtain analytical
solutions for the amide I state and for the lattice distortion, in the
form of localized sech pulses \cite{Dav1,Dav2,Brizhik83}. On the other
hand, in the more realistic discrete version numerical simulations are
the norm (see
e.g. \cite{Scott85,Law86,Law87,LK90,Scott92,CHT97,LC09,zpc16,Georgiev22}).
Here, our aim was to determine exact analytical eigenstates of the
Davydov dimer.
\\

The amount of energy released in the hydrolysis of ATP estimated from
reactions in solution is approximately 0.42~eV, enough to excite at
most two quanta of amide I \cite{Scott85}. Thus, most studies of the
Davydov model have essentially dealt with one quantum
state. 
On the other hand, other estimates that apply detailed balance and
calculate the energy necessary for specific cellular processes, can
lead to a value of 0.63~eV, which is enough to excite three quanta of
the amide I. 
Thus, in a few works, amide I states with more than two quanta have
also been investigated \cite{Law86,Law87,KL90,LK90,Georgiev22}. As
explained in the next section, we chose the mixed quantum-classical
regime of the Davydov model in which the motion of the amide I sites
is considered classical, and deal with the problem of determining
multi-quanta states exactly within this framework (unlike in studies
such as \cite{Law86,Law87,KL90,LK90,Georgiev22}). However, because the
number of coefficients needed to specify the amide I wave function is
given by \( (n+f+1)!/\left[n! (f-1)!\right]\) \cite{se86}, where $f$
is the number of sites in the lattice and $n$ is the number of quanta,
and grows generally as $f^n$, we resort to the Davydov dimer ($f=2$).
\\



\section{The Davydov dimer}
The Hamiltonian for the Davydov dimer is
\begin{equation}
  H' = \epsilon' (a_1^\dagger a_1+a_2^\dagger a_2) + J (a_1^\dagger
  a_2+a_2^\dagger a_1) - \chi u (a_1^\dagger a_1-a_2^\dagger a_2)
  + \frac{1}{2} k u^2 + \frac{1}{2 \mu } p^2,\qquad
 \label{Ham-dim}
\end{equation}
where $a_i^\dagger$ ($a_i$) are the boson creation (annihilation)
operators for an amide I vibration in site $i$, the $u$ is the
displacement from the equilibrium value of the hydrogen bond between
the two sites, $p$ is the momentum associated with $u$, $\epsilon'$ is
the energy of the amide I at zero displacement, $J$ is the
dipole-dipole interaction between amide I vibrations at neighbouring
sites, $\chi$ is the change in amide I energy with hydrogen-bond
length, $k$ is the elasticity of the hydrogen bond and $\mu$ is the
reduced mass of the two sites.
While many authors consider the regime in which both the amide I and
the hydrogen bond displacements are treated quantum mechanically,
it has been shown that, in the thermal equilibrium regime, above 11~K
its behaviour is indistinguishable from the mixed quantum-classical
regime in which the displacements are treated classically
\cite{CHK95}. Since we are concerned with events that take place at
the (much higher) biological temperatures, in what follows the changes
in hydrogen bond lengths shall be described as c-numbers. This
constitutes the mixed quantum-classical regime mentioned in the
introduction.
\\

Setting $\epsilon = \epsilon' k /\chi^2$, $V = J k / \chi^2$, and
$U = k/\chi u$ and neglecting the kinetic energy of the lattice (the
last term in (\ref{Ham-dim})) we get the following a-dimensional
Hamiltonian $H = k/\chi^2 H'$:
\begin{equation}
H = \epsilon (a_1^\dagger a_1+a_2^\dagger a_2) + V (a_1^\dagger
a_2+a_2^\dagger a_1) - U (a_1^\dagger a_1-a_2^\dagger a_2) + \frac{1}{2} U^2
\label{Ham}
\end{equation}
For the eigenstate calculations we ignore the first and last terms
which merely shift the eigenenergies. I.e., the eigenstates depend
only on $V$ and $U$.
\\

\section{Theory}

To write the wavefunction we use a site-based basis set. Applying
standard Dirac notation \cite{se86}, for the dimer we write
interchangeably
$$
\ket{\psi(n_i,n_j)} = 
   \ket{n_i}_1\ket{n_j}_2
$$
for a state with $n_i$ bosons on site 1 and $n_j$ bosons on site 2.
Note that our basis states are orthonormal
$$
{_k}\!\braket{n_i|n_j}_{\ell} =\delta_{i,j}\delta_{k,\ell}.
$$
{\color{black} With $n$ quanta we have a  basis set with $N=n+1$ elements
  $$
  \ket{\psi(n,0)},\ket{\psi(n-1,1)},\ket{\psi(n-2,2)},
  \dots,\ket{\psi(0,n)}.
  $$ }
%
%
%

\subsection{Eigenstates for n=1}
Let us first consider the case $n=1$, in which there is only one amide
I excitation in the dimer.
%
In this case there are two basis functions, namely,
$\ket{\psi_1} =\ket{\psi(1,0)}$, representing the state in which the
amide I excitation is in site $1$ and $\ket{\psi_2} =\ket{\psi(0,1)}$,
representing the state in which the amide I excitation is in site $2$.
Other choices are possible, provided we have a complete set, and it is
simplest to choose an orthonormal set. The matrix ${\bf H}$ that we
end up with (see below) depends on this choice. With the basis set we
have chosen, we have:
$$
H \ket{\psi_1} =  V\ket{\psi_2} - U\ket{\psi_1},
$$
and
$$
H \ket{\psi_2} =  V\ket{\psi_1} + U\ket{\psi_2},
$$
so the matrix $ \overline{\bf H}$ is
$$ \overline{\bf H}=\bra{\psi_i}H\ket{\psi_j}=\left[ \begin{array}{cc}
    -U & V\\ V & +U \end{array} \right].
$$
The eigenvalues are $E = \pm \surd(U^2 + V^2)$.

Since the eigenvalues are expressed in terms of $\surd(U^2 + V^2)$,
also for $n>1$, it makes sense to parameterize $U,V$ as
$U=-R\cos(\theta), V=R\sin(\theta)$.  With this change the eigenvalues
are $E = \pm R$, and the eigenvectors are simple functions of
$\cos(\theta)$ and $\sin(\theta)$.  We will keep this parameterization
in all that follows.  We now have
\begin{equation*}
  \overline{H} =  R\sin(\theta)(a_1^\dagger a_2+a_2^\dagger a_1)
  + R\cos(\theta)(a_1^\dagger a_1-a_2^\dagger a_2),
\end{equation*}
and
$$
\overline{\bf H}=\bra{\psi_i}H\ket{\psi_j} = \left[ \begin{array}{cc}
    R\cos(\theta) & R\sin(\theta)\\ R\sin(\theta) &
    -R\cos(\theta) \end{array} \right].
$$
A simple calculation gives the eigenvalues of $\overline {\bf H}$ as
$E = \pm R$, as expected.  In the matrix $\overline{\bf H}$, $R$ is an
overall multiplicative constant, which affects the eigenvalues but not
the eigenvectors.  So, for all values of $n$, we will switch to a
modified Hamiltonian with $R=1$. For $n=1$ we get:
\begin{equation}
{\bf H}= \left[ \begin{array}{cc}
    \cos(\theta) & \sin(\theta)\\ \sin(\theta) &
    -\cos(\theta) \end{array} \right].
    \label{H-1}
\end{equation}
Given the eigenvalues of this new Hamiltonian, we can recover the
eigenvalues of the original Hamiltonian (\ref{Ham}) by multiplying by
$R$ and adding \(n \, \epsilon + \frac{1}{2} \, U^2\).

In the following calculations we will be interested in increasing
values of $n$, so it is worth a short discussion of the {\em
eigenvectors} of ${\bf H}$ even in this trivial case.
Eigenvectors are only unique up to a constant multiplying factor, and
the algebraic manipulation system Maple gives the eigenvectors of
${\bf H}$ (\ref{H-1}), as
\begin{equation*}
  \left[
    -\frac{\sin \left(\theta \right)} {\cos \left(\theta \right)\pm 1} ,
    1 \right]'
\end{equation*}
which become
\begin{equation}\label{evs1}
  \left[ \sin(\theta), \mp1 - \cos(\theta)\right]'
\end{equation}
on multiplication by the denominators.

\subsection{Eigenstates for all values of $n$}

We wish to determine the analytical expressions for the eigenvalues
and eigenvectors of the matrix ${\bf H}$ (\ref{H-1}) and of its
generalizations to higher amide I quantum numbers, $n > 1$. To that
end, we make use of the Lie Algebra,
%
${\mathfrak sl}_2(\mathbb C).$ This is a complex, three dimensional
vector space with basis $\{{\bf e},{\bf f},{\bf h}\}$ and a
$\mathbb C$-linear multiplication, $*$, given by
\[
  {\bf h}*{\bf e}=2{\bf e},\quad {\bf h}*{\bf f}=-2{\bf f},\quad {\bf
    e}*{\bf f}={\bf h}.
\]

There are representations of this algebra in which the roles of the
basis elements are played by $N\times N$ complex matrices, and the
role of the multiplication operation by matrix commutation, for
example:
\[
  {\bf e}*{\bf f}=[{\bf e},{\bf f}]={\bf e}{\bf f}-{\bf f}{\bf e},
\]
etc.; the algebraic operations on the right hand side being
straightforward matrix multiplication and addition.

The $N\times N$ matrix representations of
${\mathfrak sl}_2(\mathbb C)$ are not unique but one ``standard"
representation is:

\begin{eqnarray}
	{\bf h}&=&\sum_{i=1}^N(N-2i+1){\bf E}_{i,i} \label{mat-h}\\
	{\bf e}&=&\sum_{i=1}^{N-1}\sqrt{i(N-i)}{\bf E}_{i,i+1}\label{mat-e}\\
	{\bf f}&=&\sum_{j=1}^{N-1}\sqrt{j(N-j)}{\bf E}_{j+1,j}\label{mat-f}
\end{eqnarray}
${\bf E}_{i,j}$ being the $N\times N$ matrix with unity in the
$(i,j)^{th}$ entry and zero elsewhere. Note, in particular, that
${\bf h}$ is a diagonal matrix with entries
$\{N-1,N-3,\ldots,-N+3,-N+1\}$.
\\

As shown in section \ref{appl-eq-dV}, the The ${\bf H}$ matrices of
interest to us are all of the form
\begin{equation}
{\bf H}(\theta)=\cos\theta \, {\bf h}+\sin\theta \, ({\bf e}+{\bf f})
\label{H-theta}
\end{equation}
with $N=n+1$ and we wish to show that, in the first instance, the
eigenvalues of $H$ are constant.

It is easy to see that
\begin{eqnarray}
  \frac{\partial {\bf H}(\theta)}{\partial\theta}&=&-\sin\theta \,
  {\bf h}+\cos\theta \, ({\bf e}+{\bf f})\nonumber\\
  &=&\frac12({\bf f}-{\bf e})* {\bf H}\nonumber\\
  &=&[{\bf p}, {\bf H}].\nonumber
\end{eqnarray}
where ${\bf p}=\frac12({\bf f}-{\bf e})$ is a constant matrix.

If we write $P(\theta)=\exp \left( {\bf p}\, \theta\right)$ then
\begin{eqnarray}
   {\bf H}(\theta)&=&P(\theta){\bf H}(0)P^{-1}(\theta)\nonumber\\
           &=&P\left(\theta\right){\bf h}P^{-1}\left(\theta\right)\nonumber
\end{eqnarray}	
since ${\bf H}(0)={\bf h}$. Thus, the eigenvalues of ${\bf H}(\theta)$ are
the same as those of ${\bf h}$, namely, they are its diagonal
entries, $\{-N+1,-N+3,\ldots,N-3, N-1\}$. This proves that, in
  terms of the number $n$ of amide I excitation, the energies of each
  eigenstate are simply given by:
\begin{equation}
  E_j =  jR + n \, \epsilon + \frac{1}{2} \, U^2, \hspace*{0.8cm} {\rm for} \hspace*{0.3cm} j \hspace*{0.3cm} {\rm in}
  \hspace*{0.3cm} \{-n,-(n-2),\ldots,(n-2), n\}.\label{a-en}
\end{equation}

The corresponding eigenvectors, $V$, of ${\bf H}(\theta)$ are related to $v$,
the eigenvectors of ${\bf h}$, by
\[
  V=P\left(\theta\right)v.
\]
where the $v$'s are simply the standard basis for $\mathbb C^N$.

We can construct the $V$'s by converting the above equation into a set
of linear ordinary differential equations and solving with appropriate
initial conditions.
Thus let $v$ for $i=1,2,\ldots, N=n+1$ be the standard basis for
$\mathbb C^N$.  We solve the linear system
\begin{eqnarray}
  \frac{\partial V}{\partial \theta}&=&{\bf p}V \label{eq-dV}
\end{eqnarray}
with initial condition $V(0)^{(i)}=v^{(i)}$ for each $i$.
%
Since ${\bf e}^T={\bf f}$, we have ${\bf p}^T=-{\bf p}$ and hence
$P(\theta)^TP(\theta)=\rm{Id}$. So, $P(\theta)$ is 
a real unitary operator and if we choose $V(0)^{(i)}$ to be normalized
to unity, $V(\theta)^{(i)}$ will be unit vectors also. {\color{black}
  If $V(0)$ is real, then the solution will be real also.}  In section
\ref{appl-eq-dV}, examples of using eq.(\ref{eq-dV}) to determine
exact normalized eigenvectors are presented. On the other hand, below
we develop another, purely algebraic procedure for the same purpose.
\\

The solutions to these equations belong to the span of a finite
dimensional function space which we can describe if we know the
eigenvalues of $\bf p$; that is, if we could diagonalize $\bf p$ as a
matrix with entries $\mu_i$ for $i=1,\ldots,n$, then the functions
$\exp\mu_i\theta$ will be a basis for the solution space.
In fact it is not difficult to enlist the Lie representation theory
again here. A complex representation of ${\mathfrak sl}_2(\mathbb C)$,
in which ${\bf p}$ spans the Cartan subalgebra, is given by the basis
elements,
\begin{eqnarray}
	\tilde{\bf e}&=&\frac12({\bf h}+i({\bf e}+{\bf f})),\\
	\tilde{\bf f}&=&\frac12({\bf h}-i({\bf e}+{\bf f})),\\
	\tilde{\bf h}&=&i({\bf f}-{\bf e});
\end{eqnarray}
that is, we have the commutation relations:
\[{[}\tilde{\bf h},\tilde{\bf e}]=2\tilde{\bf e},\quad {[}\tilde{\bf
    h},\tilde{\bf f}]=-2\tilde{\bf f},\quad {[}\tilde{\bf
    e},\tilde{\bf f}]=2\tilde{\bf h}.\]

Note that $\tilde{\bf h}=2i{\bf p}$.

Since $\tilde{\bf h}$ acts diagonally on its eigenvectors, we seek a
highest weight eigenvector, $\tilde v$, such that {\color{black}
  \cite{Hu72}}:
\begin{equation}\label{HW}
  \tilde{\bf h}(\tilde v)=\tilde\lambda\tilde v,\quad
  \tilde{\bf e}(\tilde v)=0.
\end{equation}

By the general representation theory \cite{Hu72} it will be enough to
show that $\tilde\lambda=n.$ The other eigenvalues will then, by
necessity, be $n-2,n-4,\ldots,-n$.

Equation ($\ref{HW}$) implies
\begin{eqnarray}
	({\bf h}+2i{\bf f})\tilde v&=&\tilde\lambda\tilde v,\\
	({\bf h}-2i{\bf e})\tilde v&=&-\tilde\lambda\tilde v,
\end{eqnarray}
and each of these equations is respectively lower and upper
triangular, so, in particular, the eigenvalues of ${\bf h}+2i{\bf f}$
are exactly the eigenvalues of ${\bf h}.$

Given the standard form of ${\bf h}$ with ${\bf h}_{11}=n$ we have an
inductive argument: if $\tilde v_1\neq 0$ then $\tilde\lambda=n$; if
$\tilde v_1=0$ then $\tilde v_2=0$ and so on.

Hence $\tilde{\bf h}$ has eigenvalues $\{n, n-2,\ldots,-n\}$ and
${\bf p}=-\frac{i}{2}\tilde{\bf h}$ has
eigenvalues
\[
  \left\{\frac{i}{2}n,\frac{i}{2}(n-2),\ldots,-\frac{i}{2}n\right\}.
\]
Our alternative construction of the eigenvectors can be made by
identifying $\theta$-dependent raising and lowering operators, defined
below as ${\bf f}(\theta)$ and ${\bf e}(\theta)$, respectively.
The corresponding eigenvectors of ${\bf p}$, $u^j$ for $j=1,\ldots,
{\color{black} N}$,
can be calculated by applying powers of $\tilde{\bf f}$ to $\tilde
v$. These are now complex valued eigenvectors. The eigenvalue of $u^j$
is $\mu_j=-i({\color{black} N}-2j+1)/2$.

To solve the problem we need only express $v^1=(1,0,\ldots,0)$ as a
linear sum of the $u^j$, with complex coefficients:
\[v^1=\sum_{j=1}^dc_ju^j,\]
then
\begin{eqnarray}
	V^1&=&\sum_{j=1}^dc_j\exp({\bf p}\theta)u^j\nonumber\\
	&=&\sum_{j=1}^dc_j\exp(\mu_j\theta)u^j\nonumber.
\end{eqnarray}

This will be a real, unit vector.

To construct $V^2,$ etc. note that we have raising and lowering operators:
\begin{eqnarray}
  {\bf e}(\theta)&=&\exp({\bf p}\theta){\bf e}\exp(-{\bf p}\theta)\nonumber\\
                 &=&\sum_{k=0}^\infty\frac{\theta^k}{k!}\textrm{ad}_{\bf p}^k
                     {\bf e}\nonumber\\
  {\bf f}(\theta)&=&\exp({\bf p}\theta){\bf f}\exp(-{\bf p}\theta)\nonumber\\
                 &=&\sum_{k=0}^\infty\frac{\theta^k}{k!}{\textrm{ad}}_{\bf p}^k
                     {\bf f}\nonumber
\end{eqnarray}	
for which we can obtain exact, finite expressions since
\[{\textrm{ad}}_{\bf p}{\bf e}=\frac12[{\bf f}-{\bf e},{\bf e}]=
  -\frac12{\bf h},\]
\[{\textrm{ad}}^2_{\bf p}{\bf e}=-\frac14[{\bf f}-{\bf e},{\bf h}]=
  -\frac12({\bf f}+{\bf e}),\]
\[{\textrm{ad}}^3_{\bf p}{\bf e}=-\frac14[{\bf f}-{\bf e},{\bf f}+{\bf e}]=
  \frac12{\bf h},\]
etc. That is,
\[{\bf e}(\theta)=-\frac12\sin\theta\,{\bf h}+\frac12(1+\cos\theta)
  {\bf e}-\frac12(1-\cos\theta){\bf f},\]
and, similarly,
\[{\bf f}(\theta)=-\frac12\sin\theta\,{\bf h}+\frac12(1+\cos\theta)
  {\bf f}-\frac12(1-\cos\theta){\bf e}.\]

Note the eigenvectors are functions of half-angles when $n$ is even and
integral angles when $n$ is odd because ${\bf f}(\theta)$ is a function of
$\cos\theta$ and $\sin\theta$ applied to half-angle functions and integral
angle functions in each case.

Now ${\bf f}(\theta)$ applied to $V^1$ gives $V^2$ and so on but the
normalization will not be preserved. To fix this we need to divide by
a factor of $\sqrt{j(N-j)}$ when applying ${\bf f}(\theta)$ to $V^j.$

We will illustrate this approach below in addition to the method
solving (\ref{mat-h}-\ref{eq-dV}).


\section{Applications of Lie theory to multi-quanta states}
\label{appl-eq-dV}

Let us now see how (\ref{mat-h}-\ref{eq-dV}) can be used to determine
the multi-quanta eigenstates of the Davydov dimer.
%
%
%
%
%
%

%
\subsection*{n=1 Case}
Starting with $n=1$, the matrices ${\bf h}$, ${\bf f}$, ${\bf e}$, and
${\bf p}$ are:
\begin{equation*}
{\bf h} = \left[\begin{array}{cc}
1 & 0
\\
 0 & -1
          \end{array}\right],\,
{\bf f} = \left[\begin{array}{cc}
0 & 0
\\
 1 & 0
\end{array}\right],\,
{\bf e} = \left[\begin{array}{cc}
0 & 1
\\
 0 & 0
  \end{array}\right],\,
{\bf p} =
 \left[\begin{array}{cc}
0 & -\frac{1}{2}
\\
 \frac{1}{2} & 0
\end{array}\right].
\end{equation*}
and the ODEs (\ref{eq-dV}) for the components of eigenvectors are:
\begin{eqnarray*}
\frac{d}{d \theta} V_1(\theta) & = -\frac12 V_2(\theta)
\\
 \frac{d}{d \theta} V_2(\theta) & = \,\,\,\,\frac12 V_1(\theta)
\end{eqnarray*}
with general solution
\begin{eqnarray*}V_1 \!(\theta) & =
  C_1 \sin \! \left(\frac12 \theta \right)
  +C_2 \cos \! \left(\frac12 \theta\right),\\
V_2 \! \left(\theta \right) & =
-C_1 \cos \! \left(\frac12 \theta\right) +
C_2 \sin \! \left(\frac12 \theta\right).
\end{eqnarray*}
For the $E_{-1}=-1$ eigenvalue, we have $V_1(0)=0,\,V_2(0)=1$ and hence
\begin{equation*}
  C_1 = -1,\quad C_2 = 0,
\end{equation*}
giving the eigenvector
\begin{equation*}
  \left[\begin{array}{c}
          \sin \left(\frac12\theta\right)\\
            -\cos \left(\frac12 \theta\right)
\end{array}\right].
\end{equation*}
which reduces to the normalized form of (\ref{evs1})  after trigometric
substitutions.  In the $E=1$ case, we have $V_1(0)=1,\,V_2(0)=0$ and hence
\begin{equation*}
  C_1 = 0,\quad C_2 = 1,
\end{equation*}
giving
\begin{equation*}
  \left[\begin{array}{c}
          \cos \left(\frac12 \theta\right)\\
          \sin \left(\frac12\theta\right)
\end{array}\right].
\end{equation*}
orthogonal to the above and again a normalised form of (\ref{evs1}).


\subsection*{$n=2$ Case}
We have
\begin{equation*}
  {\bf H}=
\left[ \begin{array}{ccc}
       -2\cos(\theta) & \surd2 \sin(\theta)  & 0 \\
        \surd2 \sin(\theta)    & 0 & \surd2 \cos(\theta)\\
         0 &    \surd2 \sin(\theta) &  2\cos(\theta)
       \end{array}      \right].
\end{equation*}
In the Lie theory we have

\begin{eqnarray*}
  {\bf h} =& \left[\begin{array}{ccc}
2 & 0 & 0
\\
 0 & 0 & 0
\\
 0 & 0 & -2
            \end{array}\right],
  {\bf f} =
\left[\begin{array}{ccc}
0 & 0 & 0
\\
 \sqrt{2} & 0 & 0
\\
 0 & \sqrt{2} & 0
      \end{array}\right],
    \\
  {\bf e} = & \left[\begin{array}{ccc}
0 & \sqrt{2} & 0
\\
 0 & 0 & \sqrt{2}
\\
 0 & 0 & 0
\end{array}\right],
{\bf p} = \left[\begin{array}{ccc}
0 & -\frac1{\sqrt{2}} & 0
\\
 \frac1{\sqrt{2}} & 0 & -\frac1{\sqrt{2}}
\\
 0 & \frac1{\sqrt{2}} & 0
\end{array}\right].
\end{eqnarray*}

The ODEs for components of eigenvectors are
\begin{eqnarray*}
\frac{d}{d \theta} V_1(\theta) & = -\frac{1}{\sqrt2} V_2(\theta)
\\
  \frac{d}{d \theta} V_2(\theta) & = \frac1{\sqrt2} V_1(\theta)
                                   -\frac1{\sqrt2} V_3(\theta) \\
  \frac{d}{d \theta} V_3(\theta) & = \frac{1}{\sqrt2} V_2(\theta)
\end{eqnarray*}
with general solution
\begin{eqnarray*}
V_1 \! \left(\theta \right) = &
  -\frac1{\sqrt2}\sin \! \left(\theta \right)\, C_3
  +\frac1{\sqrt2}\cos \! \left(\theta \right)\, C_2
  +C_1,\\
V_2 \! \left(\theta \right) = &
  C_2 \sin \! \left(\theta \right)
  +C_3 \cos \! \left(\theta \right),\\
V_3 \! \left(\theta \right) = &
   \frac1{\sqrt2} \sin \! \left(\theta \right)\,C_3
  -\frac1{\sqrt2} \cos \! \left(\theta \right) \, C_2
   +C_1.
\end{eqnarray*}
For the $E=-2,0,+2$ eigenvalues, we find in turn
\begin{eqnarray*}
  C_1 &= \frac12, C_2 = \frac1{\surd{2}}, C_3 = 0,\\
  C_1 &= 0, C_2 = 0, C_3 = 1, \\
  C_1 &= \frac12,  C_2 = -\frac1{\surd{2}}, C_3 = 0.
\end{eqnarray*}

\subsection*{$n=3$ Case}
We have
\begin{equation*}
  {\bf H}=\left[\begin{array}{cccc}
        -3 \cos \! \left(\theta \right) & \sin \!
\left(\theta \right) \sqrt{3} & 0 & 0 \\
\sin \! \left(\theta \right) \sqrt{3} & -\cos \!
\left(\theta \right) & 2 \sin \! \left(\theta \right) & 0 \\
 0 & 2 \sin \! \left(\theta \right) & \cos \! \left(\theta \right)
& \sin \! \left(\theta \right) \sqrt{3} \\
                  0 & 0 & \sin \! \left(\theta \right) \sqrt{3} & 3 \cos \!
 \left(\theta \right) \end{array}      \right].
\end{equation*}
with eigenvalues $E = -3,-1,+1,+3$.

In the Lie theory we have
\begin{equation*}
  {\bf h} =\left[\begin{array}{cccc}
3 & 0 & 0 & 0
\\
 0 & 1 & 0 & 0
\\
 0 & 0 & -1 & 0
\\
 0 & 0 & 0 & -3
          \end{array}\right],\,
        {\bf e} = \left[\begin{array}{cccc}
0 & \sqrt{3} & 0 & 0
\\
 0 & 0 & 2 & 0
\\
 0 & 0 & 0 & \sqrt{3}
\\
 0 & 0 & 0 & 0
\end{array}\right],\,
\end{equation*}
\begin{equation*}
  {\bf f} =\left[\begin{array}{cccc}
0 & 0 & 0 & 0
\\
 \sqrt{3} & 0 & 0 & 0
\\
 0 & 2 & 0 & 0
\\
 0 & 0 & \sqrt{3} & 0
\end{array}\right],\,
{\bf p} =\left[\begin{array}{cccc}
0 & -\frac12{\sqrt{3}} & 0 & 0
\\
 \frac12{\sqrt{3}} & 0 & -1 & 0
\\
 0 & 1 & 0 & -\frac12{\sqrt{3}}
\\
 0 & 0 & \frac12{\sqrt{3}} & 0
\end{array}\right].
\end{equation*}
\begin{eqnarray*}
  \frac{d}{d \theta} V_1 \! \left(\theta \right) =&
 -\frac12{\sqrt{3}}\, V_2 \left(\theta \right) ,
\\
  \frac{d}{d \theta}V_2 \! \left(\theta \right) =& \frac12{\sqrt{3}}\,
  V_1 \left(\theta \right)-V_3 \! \left(\theta \right),
\\
  \frac{d}{d \theta}V_3 \! \left(\theta \right) =&
 V_2 \! \left(\theta \right)-\frac12 \sqrt{3}\, V_4 \left(\theta \right),
\\
  \frac{d}{d \theta}V_4 \! \left(\theta \right) =& \frac12\sqrt{3}\,
   V_3 \left(\theta \right) .
\end{eqnarray*}
with solution
\begin{eqnarray*}
  V_1(\theta)  = & C_1 \sin\left(\frac32 \theta  \right)
                   + C_2 \cos\left(\frac32 \theta  \right)
                   + C_3 \sin\left(\frac12\theta \right)
                   + C_4 \cos\left(\frac12\theta \right), \\
  V_2(\theta) = &-\frac1{\sqrt{3}}\left(3 C_1 \cos\left(\frac32 \theta \right)
                  - 3 C_2 \sin\left(\frac32 \theta \right)
                  + C_3 \cos\left(\frac12\theta \right)
                  - C_4 \sin\left(\frac12\theta \right)\right) , \\
  V_3(\theta) = &-\frac{1}{\sqrt{3}} \left(3 C_2 \cos\left(\frac32 \theta \right)
                  + 3 C_1 \sin\left(\frac32 \theta \right)
                  - C_4 \cos\left(\frac12\theta \right)
                  - C_3 \sin\left(\frac12\theta \right)\right), \\
  V_4(\theta) = &C_1 \cos\left(\frac32 \theta \right)
                  - C_2 \sin\left(\frac32 \theta \right)
                  - C_3 \cos\left(\frac12\theta \right)
                  + C_4\sin\left(\frac12\theta \right).
\end{eqnarray*}
For the eigenvalues $E_i$ we have sequentially
\begin{eqnarray*}
C_1 &= 0, C_2 = \frac{3}{4}, C_3 = 0,C_4 = \frac{1}{4},\\
C_1 &= -\frac{\sqrt{3}}{4}, C_2 = 0, C_3 = -\frac{\sqrt{3}}{4}, C_4 = 0,\\
  C_1 &= 0, C_2 = \frac{\sqrt{3}}{4}, C_3 = 0, C_4 = -\frac{\sqrt{3}}{4},\\
C_1 &= -{\frac{3}{4}}, C_2 = 0, C_3 = {\frac{1}{4}}, C_4 = 0,
\end{eqnarray*}
giving the eigenvectors
\begin{eqnarray*}
\left[\begin{array}{c}
        \frac{3}{4} \cos \left(\frac{\theta}{2}\right)
        +\frac{1}{4} \cos \left(\frac{3 \theta}{2}\right)
\\
        \frac{\sqrt{3}}{4} \left(\sin \left(\frac{\theta}{2}\right)
        +\sin \left(\frac{3 \theta}{2}\right)\right)
\\
        \frac{ \sqrt{3}}{4} \left(\cos \left(\frac{\theta}{2}\right)
        -\cos \left(\frac{3 \theta}{2}\right)\right)
\\
        \frac{3}{4} \sin \left(\frac{\theta}{2}\right)
        -\frac{1}{4}\sin \left(\frac{3 \theta}{2}\right)
      \end{array}\right],
  \left[\begin{array}{c}
          -\frac{ \sqrt{3}}{4}\left(\sin \left(\frac{\theta}{2}\right)
          +\sin \left(\frac{3 \theta}{2}\right)\right)
\\
          \frac{1}{4} \cos \left(\frac{\theta}{2}\right)
          +\frac{3}{4} \cos \left(\frac{3 \theta}{2}\right)
\\
          -\frac{1}{4}\sin \left(\frac{\theta}{2}\right)
          +\frac{3}{4} \sin \left(\frac{3 \theta}{2}\right)
\\
          \frac{ \sqrt{3}}{4}\left(\cos \left(\frac{\theta}{2}\right)
          -\cos \left(\frac{3 \theta}{2}\right)\right)
\end{array}\right],
\end{eqnarray*}
\begin{eqnarray*}
  \left[\begin{array}{c}
          \frac{\sqrt{3}}{4} \left(\cos \left(\frac{\theta}{2}\right)
          -\cos \left(\frac{3 \theta}{2}\right)\right)
\\
          \frac{1}{4}\sin \left(\frac{\theta}{2}\right)
          -\frac{3}{4} \sin \left(\frac{3 \theta}{2}\right)
\\
          \frac{1}{4}\cos \left(\frac{\theta}{2}\right)
          +\frac{3}{4} \cos \left(\frac{3 \theta}{2}\right)
\\
          \frac{ \sqrt{3}}{4}\left(\sin \left(\frac{\theta}{2}\right)
          +\sin \left(\frac{3 \theta}{2}\right)\right)
\end{array}\right],
\left[\begin{array}{c}
        -\frac{3}{4} \sin \left(\frac{\theta}{2}\right)
        +\frac{1}{4}\sin \left(\frac{3 \theta}{2}\right)
\\
        \frac{ \sqrt{3}}{4} \left(\cos \left(\frac{\theta}{2}\right)
        -\cos \left(\frac{3 \theta}{2}\right)\right)
\\
        -\frac{ \sqrt{3}}{4} \left(\sin \left(\frac{\theta}{2}\right)
        +\sin \left(\frac{3 \theta}{2}\right)\right)
\\
        \frac{3}{4} \cos \left(\frac{\theta}{2}\right)
        +\frac{1}{4} \cos \left(\frac{3 \theta}{2}\right)
\end{array}\right].
  \end{eqnarray*}
  Note that the elements of these are now {\em linear} in
  $\sin$ and $\cos$ of $\frac{\theta}{2}$ and $\frac{3\theta}{2}$.

\subsection*{$n=4$ Case}

  \begin{equation*}
  {\bf H}=
\left[\begin{array}{ccccc}
4 \cos \! \left(\theta \right) & 2 \sin \! \left(\theta \right) & 0 & 0 & 0
\\
        2 \sin \! \left(\theta \right) & 2 \cos \! \left(\theta \right)
       & \sin \! \left(\theta \right) \sqrt{6} & 0 & 0
\\
        0 & \sin \! \left(\theta \right) \sqrt{6} & 0 &
      \sin \! \left(\theta \right) \sqrt{6} & 0
\\
        0 & 0 & \sin \! \left(\theta \right) \sqrt{6} &
       -2 \cos \! \left(\theta \right) & 2 \sin \! \left(\theta \right)
\\
 0 & 0 & 0 & 2 \sin \! \left(\theta \right) & -4 \cos \! \left(\theta \right)
\end{array}\right]
\end{equation*}
with eigenvalues $E =  -4,-2,0,2,4$.  
In the Lie theory we have
\begin{eqnarray*}
  {\bf h} =\left[\begin{array}{ccccc}
4 & 0 & 0 & 0 & 0
\\
 0 & 2 & 0 & 0 & 0
\\
 0 & 0 & 0 & 0 & 0
\\
 0 & 0 & 0 & -2 & 0
\\
 0 & 0 & 0 & 0 & -4
\end{array}\right], {\bf e} =\left[\begin{array}{ccccc}
0 & 2 & 0 & 0 & 0
\\
 0 & 0 & \sqrt{6} & 0 & 0
\\
 0 & 0 & 0 & \sqrt{6} & 0
\\
 0 & 0 & 0 & 0 & 2
\\
 0 & 0 & 0 & 0 & 0
                             \end{array}\right],\\
                           {\bf f} =
\left[\begin{array}{ccccc}
0 & 0 & 0 & 0 & 0
\\
 2 & 0 & 0 & 0 & 0
\\
 0 & \sqrt{6} & 0 & 0 & 0
\\
 0 & 0 & \sqrt{6} & 0 & 0
\\
 0 & 0 & 0 & 2 & 0
\end{array}\right],
  {\bf p} =\left[\begin{array}{ccccc}
0 & -1 & 0 & 0 & 0
\\
 1 & 0 & -\frac{\sqrt{6}}{2} & 0 & 0
\\
 0 & \frac{\sqrt{6}}{2} & 0 & -\frac{\sqrt{6}}{2} & 0
\\
 0 & 0 & \frac{\sqrt{6}}{2} & 0 & -1
\\
 0 & 0 & 0 & 1 & 0
\end{array}\right]
\end{eqnarray*}
\begin{eqnarray*}
  \frac{d}{d \theta} V_1 \! \left(\theta \right) =& - V_2 \left(\theta \right) ,
  \\
  \frac{d}{d \theta}V_2 \! \left(\theta \right) = &
   V_1 \left(\theta \right)  -\sqrt{\frac32} V_3 \! \left(\theta \right),
  \\
  \frac{d}{d \theta}V_3 \! \left(\theta \right) =
      & \sqrt{\frac32}V_2 \! \left(\theta \right)
     - \sqrt{\frac32}\, V_4 \left(\theta \right),
  \\
  \frac{d}{d \theta}V_4 \! \left(\theta \right) =
              &\frac12\sqrt{3}\, V_3 \left(\theta \right)- V_5(\theta),\\
  \frac{d}{d \theta}V_5 \! \left(\theta \right)   =& V_4  \left(\theta \right).
\end{eqnarray*}
general solution is
\begin{eqnarray*}
V_1 \! \left(\theta \right) = &
                                C_1 +C_2 \sin \! \left(2 \theta \right)
                                +C_3 \cos \! \left(2 \theta \right)
                                +C_4 \sin \! \left(\theta \right)
                                +C_5 \cos \! \left(\theta \right)
,\\ V_2 \! \left(\theta \right) = &
                                    -2 C_2 \cos \! \left(2 \theta \right)
                                    +2 C_3 \sin \! \left(2 \theta \right)
                                    -C_4 \cos \! \left(\theta \right)
                                    +C_5 \sin \! \left(\theta \right)
,\\ V_3 \! \left(\theta \right) = &
                -\frac{\sqrt{6}}{3} \left(3 C_2 \sin \! \left(2 \theta \right)
                                    +3 C_3 \cos \! \left(2 \theta \right)
                                    -C_1 \right)
,\\ V_4 \! \left(\theta \right) = &
                                    2 C_2 \cos \! \left(2 \theta \right)
                                    -2 C_3 \sin \! \left(2 \theta \right)
                                    -C_4 \cos \! \left(\theta \right)
                                    +C_5 \sin \! \left(\theta \right)
,\\ V_5 \! \left(\theta \right) = &
                                    C_2 \sin \! \left(2 \theta \right)
                                    +C_3 \cos \! \left(2 \theta \right)
                                    -C_4 \sin \! \left(\theta \right)
                                    -C_5 \cos \! \left(\theta \right)+C_1
\end{eqnarray*}
For the eigenvectors corresponding to $E_1,\dots,E_5$, we have respectively
\begin{eqnarray*}
C_1 & = 3/8,\quad C_2 = 0,\quad C_3 = 1/8,\quad C_4 = 0,\quad C_5 = 1/2,\\
  C_1 & = 0,\quad C_2 = 1/4,\quad C_3 = 0,\quad C_4 = -1/2,\quad C_5 = 0,\\
  C_1 & = \sqrt6/8,\quad C_2 = 0,\quad C_3 = -\sqrt6/8,\quad C_4 = 0,
        \quad C_5 = 0,\\
  C_1 & = 0,\quad C_2 = 1/4,\quad C_3 = 0,\quad C_4 = -1/2,\quad C_5 = 0,\\
  C_1 & = 3/8,\quad C_2 = 0,\quad C_3 = 1/8,\quad C_4 = 0,\quad C_5 = -1/2.
\end{eqnarray*}
So the eigenvectors are
\begin{equation*}
  \left[\begin{array}{c} 
          \frac{3}{8}+\frac18{\cos \left(2 \theta \right)}
          +\frac12{\cos \left(\theta \right)}
          \\
          \frac14{\sin \left(2 \theta \right)}
          +\frac12{\sin \left(\theta \right)}
          \\
          \frac{\sqrt{6}}{8} \left(1
          - \cos \left(2 \theta \right)\right)
          \\
          -\frac14{\sin \left(2 \theta \right)}
          +\frac12{\sin \left(\theta \right)}
          \\
          \frac{3}{8}+\frac18{\cos \left(2 \theta \right)}
          -\frac12{\sin \left(\theta \right)}
        \end{array}\right],
      \left[\begin{array}{c} 
-\frac12{\sin \left(\theta \right)}-\frac14{\sin \left(2 \theta \right)}
\\
 \frac12{\cos \left(\theta \right)}+\frac12{\cos \left(2 \theta \right)}
\\
 \frac{\sqrt{6}}{4}\sin \left(2 \theta \right)
\\
 \frac12{\cos \left(\theta \right)}-\frac12{\cos \left(2 \theta \right)}
\\
 \frac12{\sin \left(\theta \right)}-\frac14{\sin \left(2 \theta \right)}
\end{array}\right],
\end{equation*}
\begin{equation*}
\left[\begin{array}{c} 
\frac{\sqrt{6}}{8}\, \left(1-\cos \left(2 \theta \right)\right)
        \\
 -\frac{\sqrt{6}}{4} \sin \left(2 \theta \right)
\\
 \frac{1}{4}\left(3 \cos \left(2 \theta \right)+1\right)
\\
 \frac{ \sqrt{6}}{4} \sin \left(2 \theta \right)
\\
 \frac{\sqrt{6}}{8} \, \left(1-\cos \left(2 \theta \right)\right)
\end{array}\right],
\\
\left[\begin{array}{c} 
\frac14\sin \left(2 \theta \right) -\frac12\sin \left(\theta \right)
\\
 -\frac12\cos \left(2 \theta \right) +\frac12\cos \left(\theta \right)
 -\frac{\sqrt{6}}{4}\sin \left(2 \theta \right)
\\
 \frac12\cos \left(2 \theta \right)+\frac12\cos \left(\theta \right)
\\
 \frac14\sin \left(2 \theta \right) +\frac12\sin \left(\theta \right)
      \end{array}\right],\\
  \end{equation*}
\begin{equation*}
    \left[\begin{array}{c} 
            \frac{3}{8}+\frac18\cos \left(2 \theta \right)
            -\frac{1}{2}\cos \left(\theta \right)
\\
            \frac14\sin \left(2 \theta \right)
            -\frac12\sin \left(\theta \right)
\\
           \frac{\surd6}{8} (1-\cos \left(2 \theta\right))
\\
            -\frac14\sin \left(2 \theta \right)
            -\frac12\sin \left(\theta \right)
\\
            \frac{3}{8}+\frac18\cos \left(2 \theta \right)
            +\frac12\cos \left(\theta \right)
\end{array}\right]
\end{equation*}

\section*{$n=5$ Case} {\color{black}The eigenvalues are
  $E = -5,-3,-1,1,3,5$}.  The eigenvector calculations go through as
before, with the components of the eigenvectors given as linear sums
of $\cos$ and $\sin$ of $\frac{\theta}2,\frac{3\theta}2$, and
$\frac{5\theta}2$.  For example the eigenvector
  corresponding to $E=5$ is
$$
\left[\begin{array}{c}
        \frac{5}{16}  \cos \left(\frac{3 \theta}{2}\right)
        +\frac{5}{8} \cos \left(\frac{\theta}{2}\right)
        +\frac{1}{16} \cos \left(\frac{5 \theta}{2}\right)
\\
        \frac{3 \sqrt{5}}{16}\, \sin \left(\frac{3 \theta}{2}\right)
        +\frac{\sqrt{5}}{8}\, \sin \left(\frac{\theta}{2}\right)
        +\frac{\sqrt{5}}{16}\, \sin \left(\frac{5 \theta}{2}\right)
\\
        -\frac{\sqrt{10}}{16}\, \cos \left(\frac{3 \theta}{2}\right)
        +\frac{\sqrt{10}}{8}\, \cos \left(\frac{\theta}{2}\right)
        -\frac{\sqrt{10}}{16}\, \cos \left(\frac{5 \theta}{2}\right)
\\
        \frac{\sqrt{10}}{16}\, \sin \left(\frac{3 \theta}{2}\right)
        +\frac{\sqrt{10}}{8}\, \sin \left(\frac{\theta}{2}\right)
        -\frac{\sqrt{10}}{16} \, \sin \left(\frac{5 \theta}{2}\right)
\\
        -\frac{3 \sqrt{5}}{16}\, \cos \left(\frac{3 \theta}{2}\right)
        +\frac{\sqrt{5}}{8}\, \cos \left(\frac{\theta}{2}\right)
        +\frac{\sqrt{5}}{16}\, \cos \left(\frac{5 \theta}{2}\right)
\\
        -\frac{5}{16} \sin \left(\frac{3 \theta}{2}\right)
        +\frac{5}{8} \sin \left(\frac{\theta}{2}\right)
        +\frac{1}{16}\sin \left(\frac{5 \theta}{2}\right )
\end{array}\right].
$$

\section*{$n=6$ Case}
The calculations go through as before, with the components of the
eigenvectors given as linear sums of $\cos$ and $\sin$ of
$0, \theta, 2\theta$, and $3\theta$.  {\color{black} For example the
  eigenvector corresponding to $E=6$ is
$$
\left[\begin{array}{c}
        \frac{1}{32}\cos \left(3 \theta \right)
        +\frac{3}{16} \cos \left(2 \theta \right)
        +\frac{15}{32} \cos \left(\theta \right) +\frac{5}{16}
\\
        \frac{5\sqrt{6}}{32}\sin \left(\theta \right)
        +\frac{\sqrt{6}}{8} \sin \left(2 \theta \right)
        +\frac{\sqrt{6}}{32} \sin \left(3 \theta \right)
\\
        -\frac{\sqrt{15}}{16}\,  \cos \left(2 \theta \right)
        -\frac{\sqrt{15}}{32} \cos \left(3 \theta \right)
        +\frac{15}{16} -\frac{\sqrt{15}}{32} \cos \left(\theta \right)
\\
        -\frac{2\sqrt{5}}{32} \sin \left(3 \theta \right)
        +\frac{6\sqrt{5}}{32}  \sin \left(\theta \right)
\\
        -\frac{\sqrt{15}}{16}\, \cos \! \left(2 \theta \right)
        +\frac{\sqrt{15}}{32}\, \cos \! \left(3 \theta \right)
        -\frac{\sqrt{15}}{32}\, \cos \! \left(\theta \right)
\\
        \frac{5 \sqrt{6}}{32}\, \sin \! \left(\theta \right)
        -\frac{\sqrt{6}}{8}\, \sin \! \left(2 \theta \right)
        +\frac{\sqrt{6}}{32}\, \sin \! \left(3 \theta \right)
\\
        \frac{5}{16} -\frac{1}{32}\cos \left(3 \theta \right)
        +\frac{3}{16} \cos \left(2 \theta \right)
        -\frac{15}{32} \cos \left(\theta \right)
\end{array}\right].
$$
}


\section{Determination of multi-quanta eigenstates
  using purely algebraic methods.}
\label{appl-Lie-theory}
In what follows we are using superfixes to denote distinct
eigenvectors, e.g. $u^i$, $v^i$ and $V^i$, the last being the
$\theta$-dependent eigenvectors of $H(\theta)$. {\color{black} We
  consider only the cases $n=1,2$ to illustrate the technique, other
  cases follow easily along the same lines.}
\subsection*{$n=1$}
\[{\bf e}(\theta)=\left(\begin{array}{cc}-\frac12\sin\theta &
      \frac12(1+\cos\theta)\\-\frac12(1-\cos\theta) &
      \frac12\sin\theta\end{array}\right),\quad {\bf
    f}(\theta)=\left(\begin{array}{cc}-\frac12\sin\theta &
      -\frac12(1-\cos\theta)\\\frac12(1+\cos\theta) &
      \frac12\sin\theta\end{array}\right).\]

\[\tilde{\bf
    f}=\frac12\left(\begin{array}{cc}1&-i\\-i&-1\end{array}\right).\]

\[u^1=\left(\begin{array}{c}1\\i\end{array}\right),\quad u^2=
  \tilde{\bf f}u^1=\left(\begin{array}{c}1\\-i\end{array}\right).\]

Solve
\[v^1=\left(\begin{array}{c}1\\0\end{array}\right)=
  c_1\left(\begin{array}{c}1\\i\end{array}\right)+
  c_2\left(\begin{array}{c}1\\-i\end{array}\right)\]
for $c_1=c_2=\frac12.$

Then
\[V^1=\frac12\left(\begin{array}{c}1\\i\end{array}\right)
  e^{-\frac{i\theta}{2}}+\frac12\left(\begin{array}{c}1\\
           -i\end{array}\right)e^{\frac{i\theta}{2}}=
       \left(\begin{array}{c}\cos\frac{\theta}{2}\\
               \sin\frac{\theta}{2}\end{array}\right).\]
  Finally
  \[V^2={\bf f}(\theta)v^1(\theta)=\left(\begin{array}{c}
    -\sin\frac{\theta}{2}\\ \cos\frac{\theta}{2}\end{array}\right).\]

These $V^1$ and $V^2$ are correct and orthonormal with eigenvalues
$-1$ and $1$ respectively.

\subsection*{$n=2$}
\[{\bf e}(\theta)=\left(\begin{array}{ccc}-\sin\theta &
                                                        \frac{1}{\sqrt{2}}(1+\cos\theta) & 0\\
                          -\frac{1}{\sqrt{2}}(1-\cos\theta) & 0 &
                                                                  \frac{1}{\sqrt{2}}(1+\cos\theta)\\0&-\frac{1}{\sqrt{2}}(1-\cos\theta)&
                                                                                                                                         \sin\theta\end{array}\right),\]
\[{\bf f}(\theta)=\left(\begin{array}{ccc}-\sin\theta & -\frac{1}{\sqrt{2}}(1-\cos\theta) & 0\\\frac{1}{\sqrt{2}}(1+\cos\theta) & 0 &-\frac{1}{\sqrt{2}}(1-\cos\theta)\\0&\frac{1}{\sqrt{2}}(1+\cos\theta)&\sin\theta\end{array}\right)\]

\[\tilde{\bf f}=\frac12\left(\begin{array}{ccc}2&-i\sqrt{2}&0\\-i\sqrt{2}&0 &-i\sqrt{2}\\0&-i\sqrt{2}&-2\end{array}\right).\]

\[u^1=\left(\begin{array}{c}1\\i\sqrt{2}\\-1\end{array}\right),\quad u^2=\tilde{\bf f}u^1=\left(\begin{array}{c}2\\0\\2\end{array}\right),\quad u^3=\tilde{\bf f}u^2=\left(\begin{array}{c}2\\-2i\sqrt{2}\\-2\end{array}\right).\]

We could choose a different normalization here but it doesn't help
because it gets sorted at the next stage.

Solve
\[v^1=\left(\begin{array}{c}1\\0\\0\end{array}\right)=c_1\left(\begin{array}{c}1\\i\sqrt{2}\\-1\end{array}\right)+c_2\left(\begin{array}{c}2\\0\\2\end{array}\right)+c_3\left(\begin{array}{c}2\\-2i\sqrt{2}\\-2\end{array}\right)\]
for $c_1=c_2=\frac14,\,c_3=\frac18.$

Then
\[
  V^1 = \frac14\left(\begin{array}{c}1\\i\sqrt{2}\\
                       -1\end{array}\right)e^{-i\theta}
  +\frac14\left(\begin{array}{c}2\\0\\2\end{array}\right)
  +\frac18\left(\begin{array}{c}2\\-2i\sqrt{2}\\
 -2\end{array}\right)e^{i\theta} = \left(\begin{array}{c}\frac12(1+\cos\theta)\\
\frac{1}{\sqrt{2}}\sin\theta\\
\frac12(1-\cos\theta)\end{array}\right).\]
\[V^2=\frac{1}{\sqrt{2}}{\bf f}(\theta)v^1(\theta)
  = \left(\begin{array}{c}
-\frac{1}{\sqrt{2}}\sin\theta\\\cos\theta\\
\frac{1}{\sqrt{2}}\sin\theta\end{array}\right),\]
\[V^3=\frac{1}{\sqrt{2}}{\bf f}(\theta)v^2(\theta)=
  \left(\begin{array}{c}\frac12(1-\cos\theta)\\
          -\frac{1}{\sqrt{2}}\sin\theta\\
          \frac12(1+\cos\theta)\end{array}\right).
    \]

It is easy to check that these are orthonormal eigenvectors of
$H(\theta)$ with eigenvalues $-2,\,0$ and $2$ respectively.

\section{Conclusion}

We developed general procedures to determine exact eigenstates of
mixed quantum-classical Davydov dimer for any value of the amide I
quantum number $n$. We also determined a closed expression for the
eigenenergies in this system, namely, eq.(\ref{a-en}). In future work,
these developments will be applied to the calculation of physical
quantities such as the dependence on $n$ of time-dependent
probabilities of excitation in each site, the localization threshold,
and absorption spectra.

\section*{Acknowledgements}
L.C.\ acknowledges Portuguese national funds from FCT (Foundation for
Science and Technology) through projects UIDB/04326/2020,
UIDP/04326/2020 and LA/P/0101/2020.


\end{document}